\documentclass[a4paper,
              ]{jacow}

\makeatletter%
	\ifboolexpr{bool{xetex}}
	 {\renewcommand{\Gin@extensions}{.pdf,%
	                    .png,.jpg,.bmp,.pict,.tif,.psd,.mac,.sga,.tga,.gif,%
	                    .eps,.ps,%
	                    }}{}
\makeatother

\ifboolexpr{bool{xetex} or bool{luatex}} % test for XeTeX/LuaTeX
 {}                                      % input encoding is utf8 by default
 {\usepackage[utf8]{inputenc}}           % switch to utf8

\usepackage[USenglish]{babel}			 

\usepackage[final]{pdfpages}
\usepackage{multirow}
\usepackage{ragged2e}
\usepackage{subcaption}
\usepackage{array}

\ifboolexpr{bool{jacowbiblatex}}%
 {%
  \addbibresource{MOPMB004.bib}
 }{}
\begin{document}
	\title{Comparison of the Smith-Purcell Radiation Yield for Different Models}

	\author{ M.~S.~Malovytsia\textsuperscript{1,2}, N.~Delerue\textsuperscript{1}\thanks{delerue@lal.in2p3.fr}\\
		\textsuperscript{1}LAL, CNRS/IN2P3, Universit\'e Paris-Saclay, Univ. Paris-Sud, Orsay, France\\
		\textsuperscript{2}Kharkiv National University, Kharkov, Ukraine}

	\maketitle

	\begin{abstract}
		Smith-Purcell radiation is used in several applications including the measurement of the longitudinal profile of electron bunches. A correct reconstruction of such profile requires a good understanding of the underlying model. We have compared the leading models of Smith-Purcell radiation and shown that they are in agreement within the experimental errors. 
	\end{abstract}

%=======================================================================================================	
\section{Introduction}
	The production and measurement of sub-picosecond bunches is an important topic for modern accelerators.	
	To measure reliably the length of such short bunches with destroying them several approaches are possible:
	\begin{itemize}
	  \item
		  Electro-Optic (EO) sampling~\cite{Fitch99} uses a non linear crystal in which the bunch wakefield will induce optical changes. It requires a femtosecond laser. Its limitations due to material properties are discussed in~\cite{EOLimSteff09}.
	  \item
		  Coherent Transition Radiation (CTR)~\cite{Lai94} uses the radiation emitted when the beam crosses a thin foil. In some cases it may be difficult to discriminate the signal from CTR for other sources of radiation (e.g.: synchrotron radiation) generated further upstream.
	 \item
		  Coherent Smith-Purcell Radiation~\cite{Nguyen97} (CSPR),
		  uses a grating to induce the emission of radiation. It 
		  has the advantage of dispersing the radiation at the point of emission and therefore being more immune to background noise. It is described below.
	\end{itemize}
	
	To correctly interpret the CSPR signal and use it to reconstruct the bunch profile it is important to have a sound understanding of the theoretical model. 
	The focus of this paper is to compare several SPR Single Electron Yield (SEY) models to show that the model uncertainty does not significantly affect experimental measurement and profile reconstruction and to compare the signal behavior close from the grating (near-field zone).

%======================================================================================================
\section{Principle of Smith-Purcell Radiation}
\begin{figure}
		\includegraphics[width=0.5\textwidth]{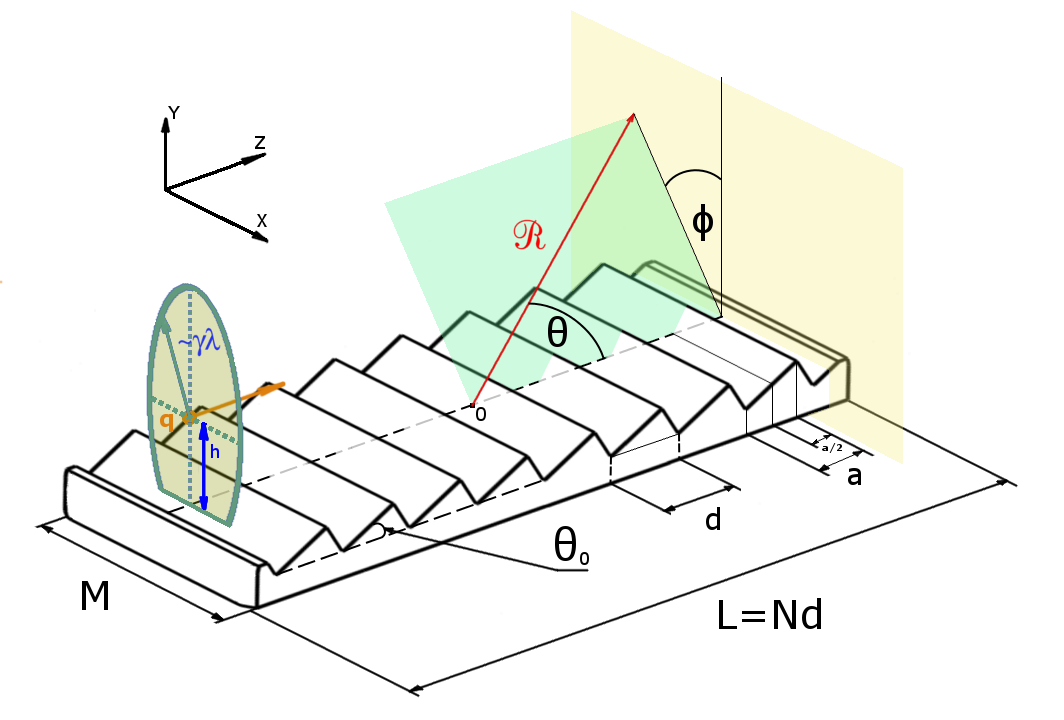}
		\caption{Definition of the variables used in this paper.}
		\label{fig:geom1}
	\end{figure}

	Smith-Purcell radiation is produced by a charged particle passing
	near a surface of a conducting periodical grating.
	In multiple papers~\cite{p020, p043, p026, p019, p039} authors considered a profile
	of the grating as a set of the periodically repeating ``$N$'' pairs of ``rising'' an ``falling''
	facets as shown on Fig.~\ref{fig:geom1}, with the period of repetition ``$d$'', a blaze angle
	``$\theta_0$'' ($\alpha_1$ in~\cite{p020, p043}), the width ``$M$'' and the length ``$L$''.
	The choice of such profile is explained in~\cite{p020}, by a possibility to do derive
	simpler analytical expressions and thus define the relation between the grating parameters and the SPR
	characteristics. It is convenient to chose the same profile for a comparison purposes.

%=======================================================================================================
\section{Single Electron Yield Models}
	The leading models to calculate the SPR Single Electron Yield (SEY) are:
	\begin{itemize}
		
	\item
		The Surface Current model~\cite{gfw}, that explains SPR through the currents that are being induced
		on the surface of the grating by a charge passing nearby. This theory has proven to be
		in a good agreement with experiments for energies from a few~MeV to 28.5~GeV~\cite{p010, p019, p043, p026}.
		\begin{equation}
			\left( \frac{dI}{d\Omega} \right)_1=2\pi q^2 \frac{L}{d^2}\frac{1}{\lambda^3}R^2\exp{\left[-\frac{2h}{\lambda_e}\right]}
		\end{equation}
		\begin{equation}
			\lambda_e=\frac{\lambda}{2\pi}\frac{\beta \gamma}{\sqrt{1+\beta^2\gamma^2\sin^2{\theta}\sin^2{\phi}}}
		\end{equation}

		Here, $q$ stands for the particle charge, $\lambda$ is the wavelength of the radiation emitted, $\lambda_e$ is an ``evanescent'' wavelength,
		$\beta, \gamma$ are the velocity of the particle and its Lorentz factor,
		$\theta, \phi$ are angles as shown on Fig.~\ref{fig:geom1}.
		$R^2$ is a grating efficiency parameter, that depends on the radiation angle and blaze angle.
	 
		Further in the paper, the results obtained with the expression for $R^2$ taken from~\cite{p021} will be called SC, and from the~\cite{gfw}, where the grating efficiency is calculated numerically, will be referred to as GFW.
	
	\item
		The Resonant Diffraction Radiation (RDR) model,
		uses equation for the diffraction radiation (DR) of an electron passing near a conductive semi-plane
		and extends it onto the case of the ``N'' periodically placed strips~\cite{p021}:
	  
		\begin{equation}\label{eq:RDR_model}
			\frac{d^2W_{RDR}}{d\omega d\Omega}=\frac{d^2W_{RDR}}{d\omega d\Omega}F_{n,cell}F_{N}
		\end{equation}
	
		Where $\frac{d^2W_{RDR}}{d\omega d\Omega}$ is a frequency distribution of the intensity of the RDR,
		$\frac{d^2W_{DR}}{d\omega d\Omega}$ is the frequency distribution of the intensity of the DR,
		$F_{N}$ is a factor corresponding to the interference from $N$ strips, 
		$F_{n,cell}$ is a term, that takes into account the interference of the DR on one strip.
	 
		For a large number of periods one can integrate Eq.~\ref{eq:RDR_model} over the frequencies and obtain an analytical expression~(see paper~\cite{p021}) for the intensity of the SPR.
	
	\item
		The model so-called Resonant Reflection Radiation (RRR) model based on the fact that a field of a moving charged particle could be described as a sum of the virtual plain waves~\cite{Mikaelian72,Haeberl94}, that will become real after scattering on the grating. The expression for the intensity of this model is given in reference~\cite{p041}.
	
		\begin{equation}\label{eq:BDR}
			\begin{split}
			&\left(
			\begin{array}{c}
			E_x^D\\
			E_z^D
			\end{array}
			\right)
			= C_1\frac{e}{\gamma\lambda}
			\displaystyle\int\limits_{-M/2}^{M/2}dX_T
			\displaystyle\int\limits_{-L/2}^{L/2}dZ_T\left(
			\begin{array}{c}
			X_T\\
			h
			\end{array}
			\right)\\
			&\times
			\chi{\left( Z_T \right)} 
			\frac{ K_1
				\left[
				\frac{2\pi}{\beta\gamma\lambda}
				\sqrt{X_T^2+h^2}
				\right]
			}
			{
				\mathcal{R}(X_T,Z_T,\theta,\phi)\sqrt{X_T^2+h^2}
			}
			\exp{
				[i \Delta \phi]
			}	  
			\end{split}
		\end{equation}
		\begin{equation}
			\begin{split}
			I&=C_1'\left( (E_x^D)^2 + (E_z^D)^2\right)
			\end{split}
		\end{equation}	
		Here,$E^D_Z,E_X^D$ are the $Z$ and $X$ components of the field on the detector, $I$ is the Intensity of the radiation, $X_T, Z_T$ are the $X, Z$ coordinates on the Fig.~\ref{fig:geom1}, $K_1$ is the modified Bessel function
		of the second order, $\chi$ equals $1$ on the grating and $0$ in the gap, $\mathcal{R}(X_T,Z_T,\theta,\phi)$ is the grating-detector distance.

		In reference~\cite{p041}, by assuming the distances from the grating to be infinite, the authors also derived the far-zone approximation of the RRR model.
	\end{itemize}
	
%=======================================================================================================
\section{Simulation of SEY for different models}
	The parameters of the SPESO at SOLEIL synchrotron and E203 at FACET at SLAC experiments~{\cite{SPESO,p046}} were used in the simulation (see table~\ref{tab:SPESO_E203}). The constant of the RRR model was calculated from the assumption, that the intensities of the SC and RRR models are equal at $\theta=90^\circ$.

	\begin{table}[!ht]
		\centering 
		\caption{The simulation parameters}
		\label{tab:SPESO_E203}
		\small
		\begin{tabular}{m{6mm} m{8mm} m{7mm} m{6mm} m{30mm} }
		 	\toprule
		 	Symb. 		& SPESO	& E203 	& Units & Description\\
		 	\midrule
		 	$\gamma$ 	& 200	& 4$\times$10\textsuperscript{4}	& 1		& The Lorentz factor\\
		 	$ d $ 		& 10	& 0.25	& mm		& The grating period			\\ 
		 	$ a $ 		& 7.5	& 0.187	& mm		& The width of one strip		\\ 
		 	$ R_0 $ 	& 310	& 220	& mm		& The distance between detector and grating\\ 
		 	$ L $ 		& 90	& 40	& mm		& The length of the grating		\\ 
		 	$ M $ 		& 20	& 20	& mm		& The width of the grating		\\ 
		 	$ h $ 		& 5		& 1		& mm		& The beam-grating separation	\\ 
		 	$ \theta_0 $& 30	& 30	& deg		& The blaze angle 				\\ 
		 	$ C_1' $ 	& 400	& 6395	& mm\textsuperscript{-3} & The normalization constant for the RRR model \\
		 	\bottomrule
		\end{tabular}
		\normalsize
	\end{table}
	\begin{figure}[!ht]
	 	 	\centering
	 	 \begin{subfigure}[t]{0.5\textwidth}
	 	 	\includegraphics[width=\textwidth]{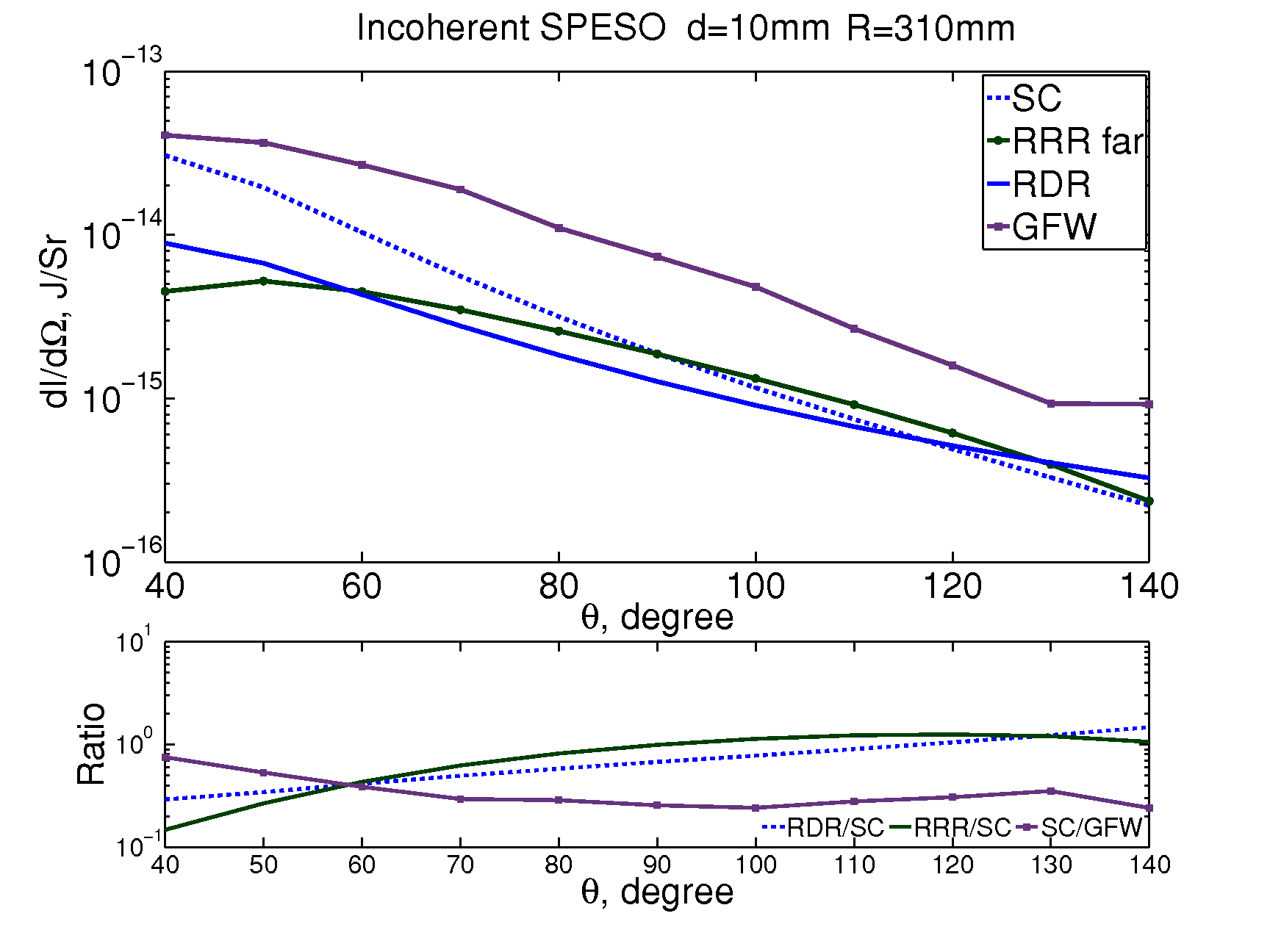}
	 	 	\caption{SPESO experiment}
	 	 	\label{fig:SPESO_RDR_SC_RRR}
	 	 \end{subfigure}
 	 	 \begin{subfigure}[t]{0.5\textwidth}
	 	 	\includegraphics[width=\textwidth]{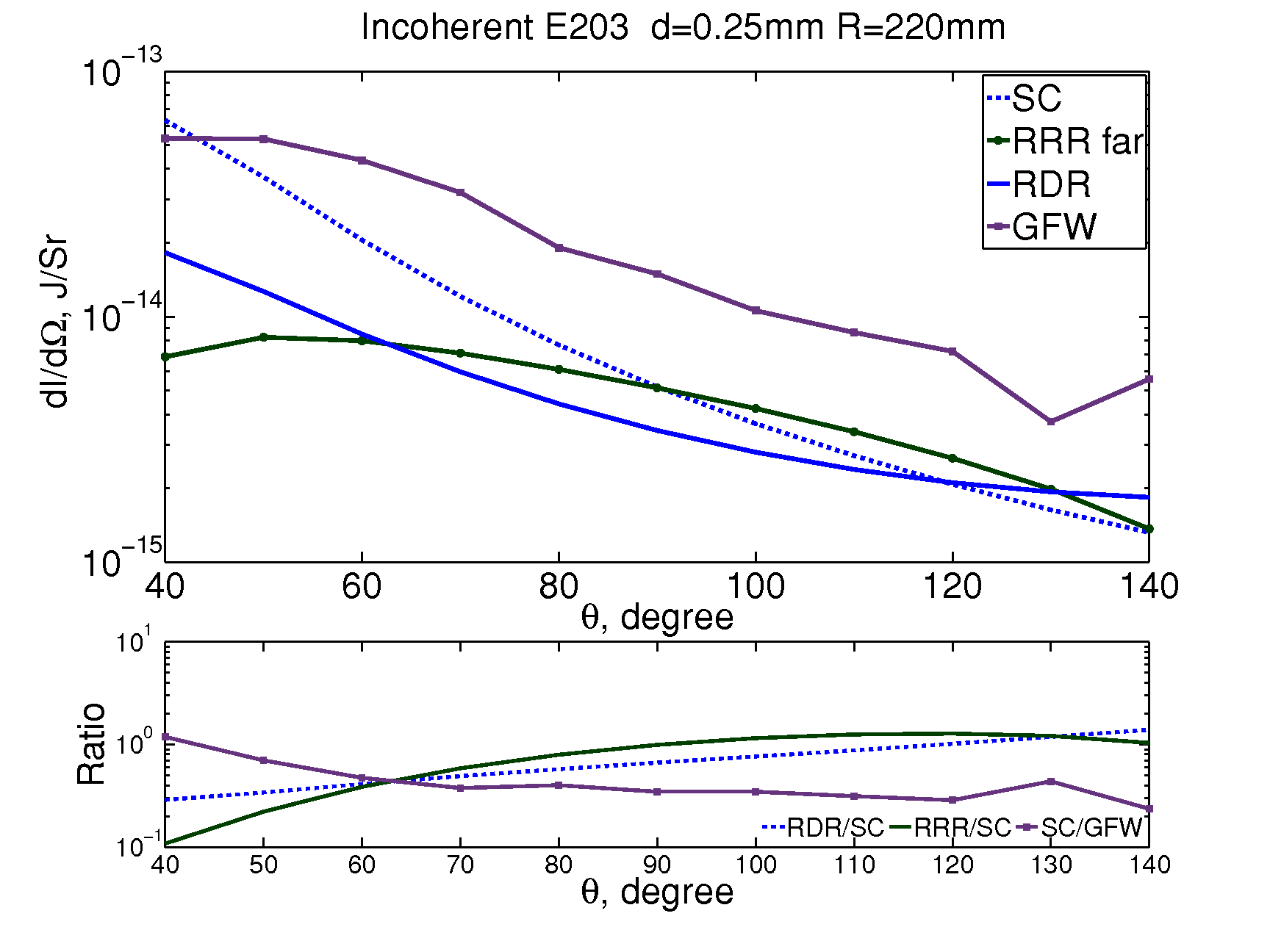}
	 	 	\caption{E203 experiment}
	 	 	\label{fig:E203_RDR_SC_RRR}
	 	 \end{subfigure}
	 	 	\caption{Calculated curves for the RDR (solid blue line), RRR (green line with circle marker), SC (blue dashed line) and GFW (purple line with square marker) models and their ratios.}
	 	 	\label{fig:theta_RDR_SC_RRR}
	\end{figure}
	 	 
	Taking into account an angular aperture of the detectors of 10$^\circ$, for each value of $\theta$ the intensity was integrated in $\phi$ over the range ${-5^\circ<\phi<5^\circ}$, in theta over the range ${\theta_i-5^\circ<\theta<\theta_i+5^\circ}$, where $\theta_i$ is the measurement angle. The calculation were done for ${40^\circ~<~\theta_i~<~140^\circ}$, with the step of $10^\circ$. 
	 
	The figures~\ref{fig:SPESO_RDR_SC_RRR},~\ref{fig:E203_RDR_SC_RRR} show the comparison of the RDR, SC, RRR in the far zone, and GFW models, and their ratio. It is seen that for the RDR, SC and RRR models the difference is not greater than a factor of 2, which is within experimental errors. The GFW model gives intensity 10 times bigger, than the RDR and SC models, which could be explained by the fact, that in GFW calculations authors take into account the width of the grating, and the grating efficiency parameter is calculated numerically, for the case of N grating facets.

%=======================================================================================================	 
\section{Conclusions}
	The SEY of the several leading models of the SPR were compared. The simulation shows that the SC and RDR models are in agreement within experimental errors. The RRR model is also close to the RDR and SC, but in our calculation one constant had to be adjusted.  GFW does a more detailed treatment of the grating profile and the simulations predict an intensity about 10 times bigger. The ratios between the models are not changing much with the parameters~(except for the observation angle). This work will allow us to estimate the error due to theoretical uncertainty when SPR is used for longitudinal profile reconstruction.
	
%----------------------------------------------------------------------------------------
%	REFERENCE LIST
%----------------------------------------------------------------------------------------


\begin{thebibliography}{99} % Use for 10-99 references
	\bibitem{Fitch99}
		M. J. Fitch	\emph{et al.},
		``Picosecond electron bunch length measurement by electro-optic detection of the wakefield'', 
		in~\textit{Proc.~PAC’99},
		New~York, USA, March-Apr.~1999,
		paper~WEA134, pp.~2181--2183.\\
	\bibitem{EOLimSteff09}
		B.~Steffen et al.,
		``Electro-optic time profile monitors for femtosecond electron bunches at the soft x-ray free-electron laser FLASH'', 
		\emph{Phys.~Rev.~ST~Accel.~Beams}, vol.~12,
		p.~032802, Mar.~2009,	\\
	\bibitem{Lai94}
		R.~Lai, U.~Happek and A.~J.~Sievers,
		``Measurement of the longitudinal asymmetry of a charged particle bunch from the coherent synchrotron or transition radiation spectrum'',
		\emph{Phys.~Rev.~E}, vol.~50,
		pp.~R4294--R4297, Dec.~1994.\\
	\bibitem{Nguyen97}
		D.~C.~Nguyen,
		``Electron Bunch Length Diagnostic With Coherent Smith-Purcell Radiation'',
		in~\emph{Proc.~PAC'97},
		Vancouver, B.C., Canada, May~1997,
		paper~97CH36167, pp.~1990--1992.	
	\bibitem{p020}
		J.~H.~Brownell and G.~Doucas.,
		``Role of the grating profile in Smith-Purcell radiation at high energies'',
		\emph{Phys.~Rev.~ST~Accel.~Beams}, vol.~8,
		p.~091301, Sept.~2005. \\ 
	\bibitem{p043}
		G.~Doucas \emph{et al.},
		``Longitudinal electron bunch profile diagnostics at 45 MeV using coherent Smith-Purcell radiation'',
		\emph{Phys.~Rev.~ST~Accel.~Beams}, vol.~9,
		p.~092801, Sept.~2006. \\
	\bibitem{p026}
		V.~Blackmore \emph{et al.},
		``First measurements of the longitudinal bunch profile of a 28.5 GeV beam using coherent Smith-Purcell radiation'',
		\emph{Phys.~Rev.~ST~Accel.~Beams}, vol.~12,
		p.~032803, Mar.~2009. \\
	\bibitem{p019}
		G.~Doucas \emph{et al.},
		``Determination of longitudinal bunch shape	by means of coherent Smith-Purcell radiation'',
		\emph{Phys.~Rev.~ST~Accel.~Beams}, vol.~5,
		p.~072802, July~2002. \\ 
	\bibitem{p039}
		V.~Blackmore \emph{et al.},
		``First observation of coherent Smith-Purcell radiation in the highly relativistic regime.''
		\emph{Nucl.~Instrum.~Methods~Phys.~Res., Sect.~B}, vol.~266, no.~17,
		pp.~3803--3810, 2008. \\
	\bibitem{gfw}
		J.~H.~Brownell, J.~Walsh, G.~Doucas,
		``Spontaneous Smith-Purcell radiation described through induced surface currents'',
		\emph{Phys.~Rev.~E} vol.~57,
		pp.~1075--1080,  Jan.~1998.\\
	\bibitem{p010}
		G.~Doucas \emph{et al.},
		``First observation of Smith-Purcell radiation from	relativistic electrons'',
		\emph{Phys.~Rev.~Lett.}, vol.~69,
		pp.~1761--1764, Sept.~1992. \\	
	\bibitem{p021}
		D.~V.~Karlovets and A.~P.~Potylitsyn.
		``Comparison of Smith-Purcell radiation models and criteria for their verification'',
		\emph{Phys.~Rev.~ST~Accel.~Beams}, vol.~9,
		p.~080701, Aug.~2006. \\
	\bibitem{Mikaelian72}
		M.~L.~Ter-Mikaelian,
		\emph{High Energy Electromagnetic Processes in Condensed Media}.
		New~York, USA:
		John~Wiley and Sons Inc, 1972. \\
	\bibitem{Haeberl94}
		O.~Haeberl\'e  \emph{et al.},
		``Calculations of Smith-Purcell radiation generated by electrons of 1\char21{}100 MeV'',
		\emph{Phys.~Rev.~E}, vol.~49,
		pp.~3340--3352, Apr.~1994. \\
	\bibitem{p041}
		D.~V.~Karlovets and A.~P.~Potylitsyn.,
		``Smith-Purcell radiation in the ``pre-wave'' zone”,
		\emph{JETP~Letters}, vol.~84, no.~9,
		pp.~489-–493, 2006. \\
	\bibitem{SPESO}
		N.~Delerue  \emph{et al.},
		`First Measurements of Coherent Smith-Purcell Radiation in the SOLEIL Linac'',
		paper~MOPMB002, these proceedings.\\
	\bibitem{p046}
		N.~Delerue \emph{et al.},
		``Electron Bunch Profile Diagnostics in the Few fs Regime using Coherent Smith-Purcell Radiation'',
		in~\textit{Proc.~IPAC’11},
		San~Sebastian, Spain, Sept.~2011,
		paper~MOP057, pp.~567--569.\\
\end{thebibliography}
\end{document}